# Interaction effects and superconductivity signatures in twisted double-bilayer WSe$_2$

*Liheng An[†], Xiangbin Cai[†], Meizhen Huang, Zefei Wu, Jiangxiazi Lin, Zhehan Ying, Ziqing Ye, Xuemeng Feng and Ning Wang[*]*

*Department of Physics, Center for Quantum Materials and Willian Mong Institute of Nano Science and Technology, the Hong Kong University of Science and Technology, Clear Water Bay, Hong Kong, China*

[†]These authors contributed equally to this work.

[*]Corresponding author: phwang@ust.hk (N. W.)

## Abstract

Twisted bilayer graphene provides a new two-dimensional platform for studying electron interaction phenomena and flat band properties such as correlated insulator transition, superconductivity and ferromagnetism at certain magic angles. Here, we present strong evidence of correlated insulator states and superconductivity signatures in p-type twisted double-bilayer WSe$_2$. Enhanced interlayer interactions are observed when the twist angle decreases to a few degrees as reflected by the high-order satellites in the electron diffraction patterns taken from the 2H/3R-stacked domains reconstructed from a conventional moiré superlattice. In contrast to twisted bilayer graphene, there is no specific magic angle for twisted WSe$_2$. The flat band properties are observed at twist angles ranging from 1 to 4 degrees. The highest superconducting transition temperature observed by transport measurement is 6 K. Our work has facilitated future study in the area of flat band related properties in twisted transition metal dichalcogenide layered structures.

The discovery of superconductivity and exotic insulating phases in twisted bilayer graphene[1] has led to an increased interest in the study of modulated flat band properties and correlation effects in two-dimensional (2D) systems. Artificially stacking one single-layer graphene with respect to another flake of single-layer graphene with a small specified rotation angle, the so-called 'magic' angle, generates a moiré pattern or moiré unit-cells, which could reshape the low energy electronic states, resulting in new flat bands and many localized states. These flat bands have been proven to host many exotic quantum phenomena such as Mott-insulator transition,[2] unconventional superconductivity and ferromagnetism.[3,4] Similar flat band properties have also been evidenced in twisted double-bilayer graphene[5,6] and in ABC stacked trilayer graphene moiré superlattices formed on hexagonal boron nitride (h-BN) sheets.[7]

To date, these unusual flat band characteristics have primarily been observed in graphene-graphene moiré superlattices created at a precisely controlled small twist angle (about or smaller than 1º) in order to enhance the electron-electron Coulomb interaction effects. The electron or hole density needed for fully filling the moiré superlattice unit cell is restricted to the level of $\sim 4 \times 10^{12}/cm^2$. Having a geometry similar to that of graphene, atomically thin semiconducting transition metal dichalcogenides (TMDCs) are potential candidate materials for fabricating 2D twisted heterostructures and exploring their transport and optical properties[8] governed by flat bands and correlation effects since this kind of layered materials have large effective masses, relatively strong electron-electron interactions[9-11] and more additional effects stemming from the lack of inversion symmetry and large spin-orbit interactions. Recent studies have indicated that ultra-flat bands and interaction effects could be realized by moiré quantum well structures in the valence band of twisted bilayer TMDCs.[12-14,28] Here, we present experimental results demonstrating the successful fabrication of twisted double-bilayer $WSe_2$ device structures with enhanced interlayer interactions as observed by atomic resolution imaging of the 2H/3R-stacked domains reconstructed from a conventional moiré superlattice. Correlated insulator states and superconductivity signatures are observed by transport measurement at cryogenic temperatures. Distinct from graphene, the flat band properties can be observed in $WSe_2$ at relatively large twist angles ranging from 1º to 4º; this offers new design opportunities for fabricating moiré quantum well heterostructures.

Figure 1a,b illustrate the device structure of the twisted double-bilayer $WSe_2$ fabricated by standard exfoliation and dry transferring techniques reported previously.[29] The twisted double-bilayer $WSe_2$ is encapsulated by h-BN sheets and controlled in a double-gating configuration for this study. Platinum (Pt) bottom electrodes[15] are used to make electrical contacts to the $WSe_2$ channels. Ohmic contacts are achieved by applying a large negative bias via the top-gate, which enables the carrier density to be tuned at the same time. The crystallographic structures of the twisted double-bilayer $WSe_2$ are investigated by high-resolution scanning transmission electron microscopy (STEM). Using the same transferring and stacking techniques, we mechanically transfer the double-bilayer $WSe_2$ flakes with desired twist angles onto lacey carbon grids for STEM study. As shown in the selected-area electron diffraction patterns in Figure 1e, the 1.1º twist angle between the two sets of hexagonal diffraction patterns is evident which is close to the designed angle of 1º with an acceptable uncertainty of ±0.1º. Theoretically, overlapping two layers of a crystalline structure at a small twist angle generates a moiré pattern as schematically depicted in Figure 1c. In this case, one should observe two sets of hexagonal diffraction patterns (the Bragg diffractions). As illustrated in the enlarged pictures of certain diffractions such as (010), (110) and (020) in Figure 1f-h, however, we observe many high-order satellites (from first to third orders) surrounding each pair of Bragg diffractions, indicating that strong interlayer interactions or interface reconstructions occur in the twisted double-bilayer $WSe_2$. According to the reciprocal spacing of the satellite diffractions, the lattice parameter of the newly formed unconventional moiré unit-cell should be 20-25 nm.

As verified by aberration-corrected annular dark-field imaging in STEM, the twisted $WSe_2$ bilayers undergo a structural reconstruction and form alternatingly distributed triangular regions as marked in blue and red in Figure 1i. The scale of the triangular pattern is consistent with the reciprocal space measurement. The atomic-resolution images (the insets in Figure 1i) reveal that 2H and 3R stacking orders of $WSe_2$ exist in the blue and red regions respectively, further verified that the structural reconstruction from a conventional moiré pattern occurs and produces well-relaxed stacking domains of more stable crystalline structures. With the AB stacking in 2H bilayer $WSe_2$, the small twist angle causes the stacking sequence of the $WSe_2$ layers to transit from the common ABAB stacking (2H structure) to ABCA stacking (3R structure) as shown in Figure 1d. The 2H and 3R domains are separated by the hexagonal network boundaries (~3 nm in width). The narrow domain boundaries have lattice misfit strain as reflected by the image dark contrast. However, the image contrasts in ABAB and ABCA areas are very uniform, indicating that lattice misfit strain has been relaxed.

The band structure of $WSe_2$ is layer number dependent. In monolayer and bilayer $WSe_2$,[16] the valence band maxima locate at K-valleys. In a few-layer $WSe_2$ with a layer number greater than three,[9] $\Gamma$–valley heavy holes dominate the carrier transport in the valence band. In a four-layer $WSe_2$ device, the electrical conductance measured at cryogenic temperatures (see Figure 3a) monotonically increases with increasing the gate voltage or carrier concentration. In twisted double-bilayer $WSe_2$, however, the conductance normally exhibits a few peaks and dips with increasing the gate voltage (see Figure 2a), indicating the existence of localization resulted from the moiré superlattice structures. As an example, Figure 2b,c illustrate the measured longitudinal resistance of a sample with a 4° twist angle at different gating configurations. By setting the top gate $V_{TG}$ to -22V (through a 25nm h-BN layer), the channel resistance measured by a four-lead configuration is about 200 ohms (see Figure 2b) and the corresponding carrier density is $9.9 \times 10^{12}/cm^2$. The hole mobility calculated by $\mu_h = \frac{Ld\sigma}{Wdn}$ has reached 2000$cm^2$/Vs, which is similar to that of the high performance devices of monolayer and bilayer $WSe_2$ reported previously.[10, 17]

For the twisted double-bilayer (totally 4 layers) device, the valence band edges locate at the $\Gamma$-valley with a two-fold degeneracy. Given that, the two peaks shown in the longitudinal resistance at $V_{TG}$ = -15.4V and -19.8V in Figure 2b, could then correspond to the correlated insulating state at $4.8 \times 10^{12}/cm^2$ and the band insulating state at $8.4 \times 10^{12}/cm^2$ respectively based on the following estimation. The moiré pattern superlattice constant is expressed by $\lambda = a/(2\sin(\frac{\theta}{2}))$, where *a* is the lattice constant of $WSe_2$. $\lambda$ is estimated to be 4.7nm for a 4° twist angle. The full filling of the moiré superlattice can be worked out as $n_0 = 2 \times 10^{-2} / \frac{\sqrt{3}}{2}\lambda^2 \approx 1.05 \times 10^{13}/cm^2$. This value is quite close to the carrier concentration calculated according to the quantum capacitance of the WSe$_2$-BN heterostructure. In principle, there are two metallic states adjacent to the correlated half filling insulating states and band insulating states. The first one locates at V$_{TG}$= -14.8V and displays metallic characteristics since the resistance measured at this gate voltage decreases with decreasing sample temperature from 300 K to 10 K. While, it does not change so much below 10 K. The second one locates between V$_{TG}$ = -16V to -19.6V with a carrier density ranging from 5.9 to 8.3×10$^{12}$/cm$^2$. It has a metallic behavior in the temperature region of 11-125 K. The measured resistance decreases gradually with decreasing sample temperature, which means that phonon scattering dominates in the electron transport behavior. From 11 K to 1.9 K, however, a quick reduction of the resistance (from 1600 to 200 ohms) occurs as shown in Figure 2d. Such a quick reduction of the channel resistance is attributed to the formation of superconducting states at the interface of twisted double-bilayer WSe$_2$. Because of the impurity-dominated scattering in WSe$_2$, the intrinsic resistance of a few-layer WSe$_2$ device normally remains as a constant value at very low temperatures and have no contribution to the observed quick reduction of the channel resistance.

Ultra-flat bands formed at the edges of the valence band in WSe$_2$ moiré superlattice have been predicted theoretically.[12-14] The band width is comparable to that of twisted graphene at the 'magic' angle. In these flat bands, kinetic energy is quenched (E$_k$ ~ 10meV) which is in the same order as the on-site Coulomb interaction U~$\frac{e^2}{\varepsilon\lambda}$, where $\varepsilon = 4$ is the dielectric constant of h-BN. For λ = 4.7nm, the Coulomb interaction is about dozens of meV. This could lead to the superconducting states at the interface of twisted double-bilayer WSe$_2$. By fixing the carrier density at $7\times 10^{12}/cm^2$, we observe that the longitudinal resistance decreases quickly when sample temperature is below 18 K (Figure 2d). This transition temperature is much higher than the highest onset transition temperature of twisted graphene. At 6 K, the channel resistance drops to 800 Ω, about half the normal metallic state resistance, indicating a transition temperature of $T_C \approx 6\ K$. This finding suggests that the interaction effects induced by twisting bilayer WSe$_2$ are substantial. At the superconducting states, the non-zero residual resistance is attributed to the structural inhomogeneity in the twisted layers of the sample. As observed by electron microscopy, the twist angle often slightly changes from area to area, resulting in micrometer-sized domains in the sample. Because our sample size is about a few micrometers, the electrodes may cross more than one domain. That could be the reason why we often observe more than one dome of the superconducting states. In this case, the measured channel region from one superconducting dome is in series connection with one or more slightly different domains and boundaries, leading to the non-zero resistance.

A small twist angle leads to the formation of a large moiré superlattice cell. For a 1° twist angle (λ ≈ 19nm), fully filling the moiré superlattice, as shown in Figure 1i, requires $n_0 = 2 \times 10^{-2} / \frac{\sqrt{3}}{2}\lambda^2 \approx 0.65 \times 10^{12}/cm^2$, an order of magnitude smaller than that of the 4° twisted sample. Similarly, by changing the top-gate voltage, we observe multiple superconducting states in the 1° twisted device (Figure 3a), which could be the result of high order flat bands. The thickness of the h-BN dielectric layer in this device is about 25nm. The separation between the first and second superconducting domes is $\Delta V \approx 1V$, corresponding to a carrier density of $0.66 \times 10^{12}/cm^2$, which is very close to the full filling state of the moiré superlattice unit cell shown in Figure 1i. The two-terminal conductance measurement shows some bumps, which correspond to the superconducting states observed in the four-terminal resistance measurement. In Figure 2d, we find that the resistance remains nearly a constant value above 5 K, and a quick resistance drop happens below 5 K. The superconductivity transition temperature is then estimated to be about 3 K, which is lower than that of the 4° twisted device. Obviously, interaction effects in this kind of small-twist-angle devices are relatively weak.

The multiple superconducting domes shown in Figure 3a are attributed to high-order flat bands. Because of structural inhomogeneity in the twisted devices, the domes do not occur periodically. In addition to the thermal effects on the superconductor states, we observe that a perpendicular magnetic field can change the superconducting states to normal metallic states. Figure 3b illustrates the changes of the longitudinal resistance as a function of applied perpendicular magnetic fields at 1.9 K. The critical magnetic field needed to break these superconducting states is about 0.5 T, which points to Zeeman effects interfering with the superconducting states. Because of the structural inhomogeneity in the twisted WSe$_2$ devices, measuring the critical current density undergoes difficulties. This is because the induced electric field via V$_{ds}$ across the inhomogeneous superconducting domains and their boundaries cannot be well defined. The multiple and inhomogeneous superconducting domes cause the complexity to accurately determine the critical current in the sample because the measured critical current displays fluctuation and variation from one dome to another. Nevertheless, the general signature of the critical current density can still be seen through two-terminal conductance measurement as shown in Figure 3c, in which the peaks of the conductance (σ) correspond to the formation of the superconducting channels or the longitudinal resistance minima positions. These superconducting features, which can be seen under a small excitation of V$_{ds}$ = 6 meV, gradually disappear under a relative large excitation. We roughly estimate the critical current to be about 40 nA or higher since the superconducting states at V$_{TG}$ = -16V disappear at an excitation V$_{ds}$ = 15meV. Further improvement on the homogeneity and stability of twisted WSe$_2$ devices is needed in order to explore the detailed mechanisms of the superconductivity and correlated insulating phases.

In conclusion, we successfully fabricated a few twisted double-bilayer WSe$_2$ devices with well-controlled twist angles. Our samples demonstrate reconstructed moiré superlattices, for which the hosted carrier densities range from $0.65 \times 10^{12}/cm^2$ to $1.05 \times 10^{13}/cm^2$. We observe superconductivity signatures with relatively higher transition temperatures than that of twisted graphene devices. With strong intrinsic electron-electron interaction effects and large effective masses, atomically thin twisted WSe$_2$ could be a new platform for studying the correlated behavior of 2D electrons.

**ACKNOWLEDGMENTS**

We thank Prof. Xi Dai from HKUST for fruitful discussions. This work is supported by the Research Grants Council of Hong Kong (Project No. 16300717 and C7036-17W). We also acknowledge the technical support from the Super-resolution Electron Microscopy facility (C6021-14E) and Raith-HKUST Nanotechnology Laboratory for the electron-beam lithography facility at MCPF.

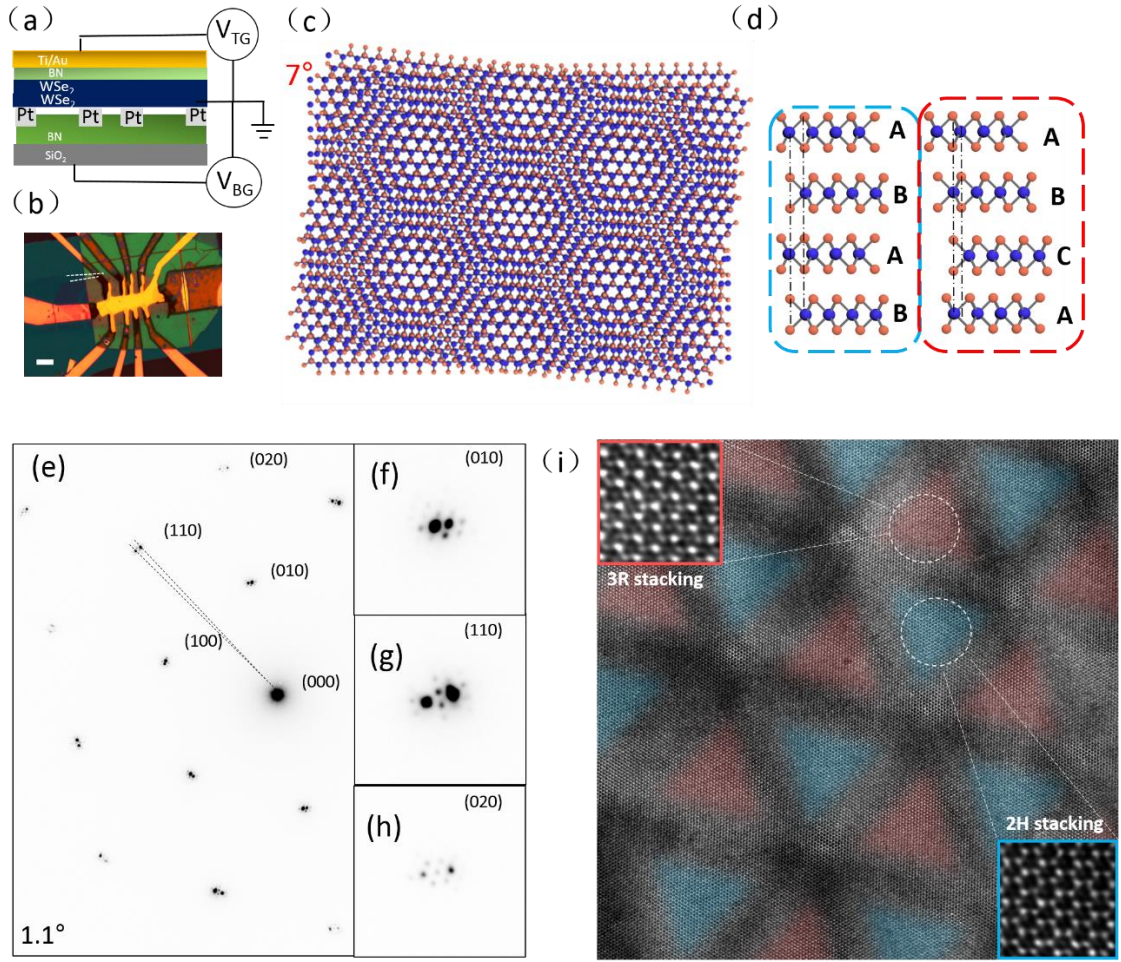

**Figure 1.** Device structure, stacking model and atomic-scale reconstruction in twist double bilayer WSe$_2$. (a) and (b) Schematic and optical images of the encapsulated twisted double-bilayer WSe$_2$. The scale bar is around 6µm. (c) A schematic moiré superlattice formed by twisted double-bilayer WSe$_2$ flakes. (d) Atomic model of 2H-stacked (ABAB) and 3R-stacked (ABCA) double-bilayer WSe$_2$ viewed along the <100> zone axis. (e) Electron diffraction pattern taken from a twisted double-bilayer WSe$_2$ with enlarged views of respective Bragg spots (f)-(h), showing the high-order satellites caused by interface reconstruction. (i) Annular dark field image of the interface-reconstructed region, in which 2H/3R-stacked domains are marked in blue/red. The zoomed-in images illustrate atomic structures.

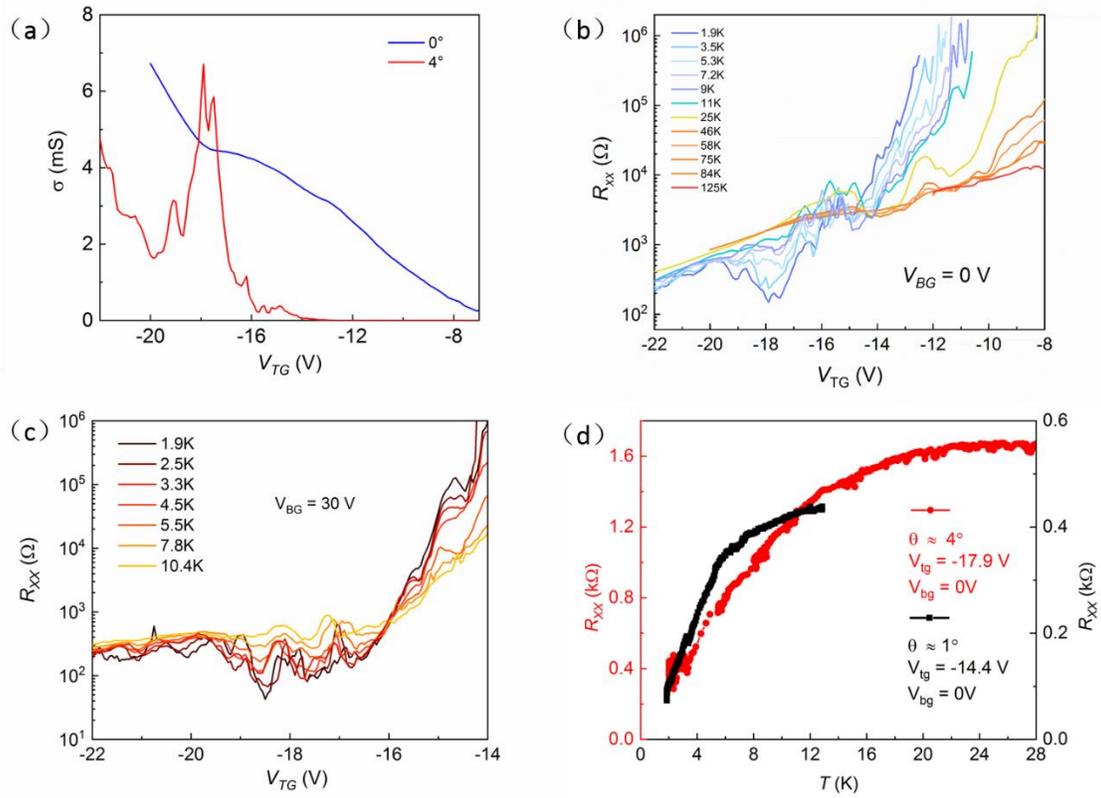

**Figure 2.** Experimental data of superconductivity obtained from twist double bilayer $WSe_2$. (a) Conductance of intrinsic four layers $WSe_2$ and twisted double-bilayer $WSe_2$ devices. (b) Longitudinal resistance plotted as a function of the top-gate voltage $V_{TG}$ at different temperatures. The excitation $V_{ds} = 1mV$ and $V_{BG} = 0$ V. (c) Longitudinal resistance plotted as a function of the top-gate voltage $V_{TG}$. Data are collected in the temperature range 1.9 - 10.4K under excitation $V_{ds} = 1mV$ and $V_{BG} = 30V$. (d) Quick resistance reduction at low temperature for the 1° and 4° twisted devices.

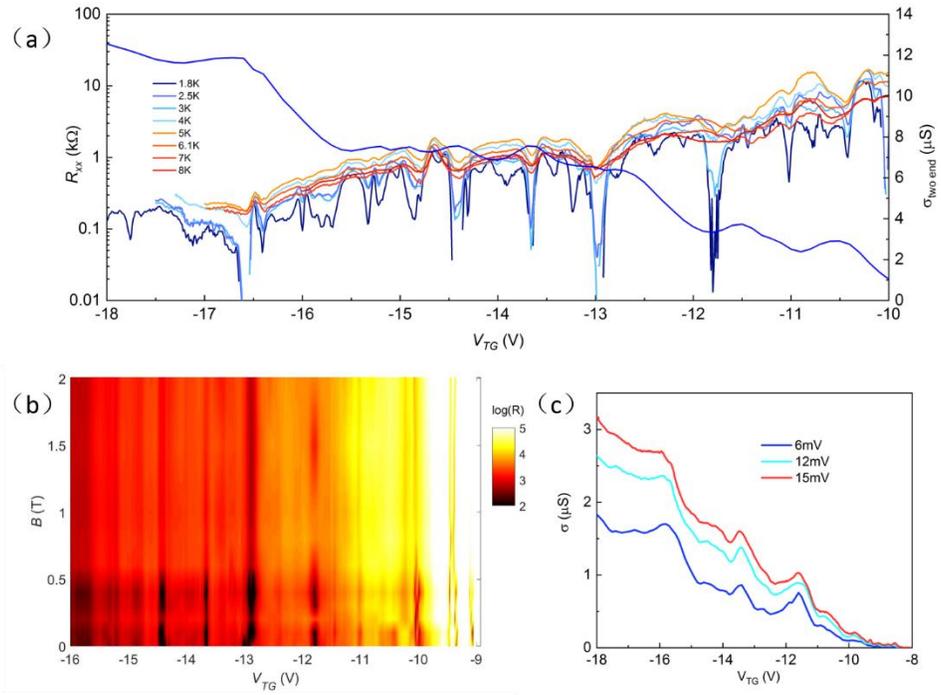

**Figure 3.** Superconductor characteristics in 1° twist double bilayer WSe$_2$. (a) Longitudinal resistance measured by four terminal configuration of the 1° twist device. The single blue line presents the conductance of the same sample measured by two terminals. (b) The longitudinal resistance plotted as a function of perpendicular magnetic fields at 1.9K. The map intensity is Log ($R_{xx}$). (c) Two-terminal conductance measured under different excitation voltages $V_{ds}$.